\documentclass[manuscript]{emulateapj}
\usepackage{hyperref}
\usepackage{epstopdf}
\newcommand{\lsim}{\ \raise -2.truept\hbox{\rlap{\hbox{$\sim$}}\raise 5.truept\hbox{$<$}\ }}
\newcommand{\gsim}{\ \raise -2.truept\hbox{\rlap{\hbox{$\sim$}}\raise 5.truept\hbox{$>$}\ }}

\begin{document}
\title{Stellar-Encounter Driven Red-Giant Star Mass-Loss in Globular Clusters}
\shorttitle{Stellar-Encounter Driven RGB Mass-Loss in GCs}
\shortauthors{Pasquato et al.}


\author{Mario Pasquato}
\affil{Department of Astronomy \& Center for Galaxy Evolution Research, Yonsei University, Seoul 120-749, Republic of Korea}
\affil{Yonsei University Observatory, Seoul 120-749, Republic of Korea}

\author{Andrea de Luca}
\affil{Laboratoire de Physique Th\'eorique, Ecole Normale Sup\'erieure 24, rue Lhomond 75231 Paris Cedex 05 - France}

\author{Gabriella Raimondo}
\affil{INAF-Osservatorio Astronomico di Teramo, Mentore Maggini s.n.c., 64100 Teramo, Italy}

\author{Roberta Carini}
\affil{INAF-Osservatorio Astronomico di Roma, Via Frascati 33, 00040, Monte Porzio Catone, Italy}

\author{Anthony Moraghan}
\affil{Department of Astronomy \& Center for Galaxy Evolution Research, Yonsei University, Seoul 120-749, Republic of Korea}
\affil{Academia Sinica Institute of Astronomy and Astrophysics, P.O. Box 23-141, Taipei 106, Taiwan}

\author{Chul Chung}
\affil{Department of Astronomy \& Center for Galaxy Evolution Research, Yonsei University, Seoul 120-749, Republic of Korea}

\author{Enzo Brocato}
\affil{INAF-Osservatorio Astronomico di Roma, Via Frascati 33, 00040, Monte Porzio Catone, Italy}

\author{Young-Wook Lee}
\affil{Department of Astronomy \& Center for Galaxy Evolution Research, Yonsei University, Seoul 120-749, Republic of Korea}

\begin{abstract}
Globular Cluster (GC) Color-Magnitude Diagrams (CMDs) are reasonably well understood in terms of standard stellar-evolution. However, there are still some open issues, such as fully accounting for the Horizontal Branch (HB) morphology in terms of chemical and dynamical parameters. Mass-loss on the Red Giant Branch (RGB) shapes the mass-distribution of the HB stars, and the color distribution in turn. The physical mechanisms driving mass-loss are still unclear, as direct observations fail to reveal a clear correlation between mass-loss rate and stellar properties. The horizontal-branch mass-distribution is further complicated by Helium-enhanced multiple stellar populations, because of differences in the evolving mass along the HB. We present a simple analytical mass-loss model, based on tidal stripping through Roche-Lobe OverFlow (RLOF) during stellar encounters. Our model naturally results in a non-gaussian mass-loss distribution, with high skewness, and contains only two free parameters. We fit it to the HB mass distribution of $4$ Galactic GCs, as obtained from fitting the CMD with Zero Age HB (ZAHB) models. The best-fit model accurately reproduces the observed mass-distribution. If confirmed on a wider sample of GCs, our results would account for the effects of dynamics in RGB mass-loss processes and provide a physically motivated procedure for synthetic CMDs of GCs. Our physical modeling of mass-loss may result in the ability to disentangle the effects of dynamics and helium-enhanced multiple-populations on the HB morphology and is instrumental in making HB morphology a probe of the dynamical state of GCs, leading to an improved understanding of their evolution.
\end{abstract}





\section{Introduction}
Globular Clusters (GCs) are the environment of choice for testing our theoretical understanding of stellar evolution.
Stellar evolution models have generally been able to reproduce the features of GC Color-Magnitude Diagrams (CMDs), yet some open issues still remain, such as that of mass-loss along the Red Giant Branch (RGB), which in turn influences the later stages of stellar evolution \citep[][]{1991A&ARv...2..249L, 1996ASPC..109..481H, 2002ApJ...571..458O, 2007ApJ...667L..85O, 2009AJ....138.1485D, 2010ApJ...718..522O}, in particular the color and temperature distribution of stars along the Horizontal Branch (HB).
This problem is also difficult to decouple from the influence of Helium enhanced multiple stellar populations, which are widespread in GCs and heavily influence HB morphology \citep[see][]{2004ApJ...605L.125B,  2004ARA&A..42..385G, 2005ApJ...621..777P, 2010A&A...517A..81G, 2010A&A...519A..60B,2013MNRAS.431.2126M}.
Synthetic CMDs are usually built assuming an arbitrary RGB mass-loss distribution such as a gaussian with adjustable variance \citep[][]{1973ApJ...184..815R, 1994ApJ...423..248L, 2000A&AS..146...91B, 2002ApJ...569..975R}.
Various mass-loss laws \citep[see][for a review]{2009Ap&SS.320..261C} have been proposed and implemented in stellar evolution prescriptions \citep[e.g.][]{1981A&A....94..175R}
with Reimers law \citep[][]{1978A&A....70..227K} remaining the most prominent.
However a full understanding of the underlying physics of wind-driven mass-loss is still elusive \citep[e.g. see][and references therein]{2009AJ....138.1485D, 2010ApJ...711L..99B} and direct observations of live mass-loss by detection of warm circumstellar dust as far infrared excess \citep[][]{2012A&A...540A..32G, 2012A&A...541C...3G, 2007ApJ...667L..85O} are feasible but difficult in crowded GC environments \citep[][]{2010ApJ...711L..99B, 2010ApJ...718..522O}. It is likely that future ALMA observations may help shed further light on this and related issues \citep[][]{2008Ap&SS.313..201O, 2009A&A...506.1277G}.

{A low-mass star would have to lose a few tenths of a solar mass on the RGB in order to be positioned on the hot end of the zero-age HB (ZAHB). Instead, a lower mass loss (i.e. of the order of a few hundredths of a solar mass) is required to produce cool ZAHB stars. There is evidence that mass loss may be due to distinct physical processes in order to populate different temperature ranges on the HB \citep[see, e.g., ][]{1998ApJ...500..311F}. Moreover, it may take place in relatively brief events and may be weakly dependent on stellar properties \citep[][]{2002ApJ...571..458O, 2007ApJ...667L..85O, 2009MNRAS.394..831M}, as would be expected if the mass-loss mechanisms were not fully intrinsic to the star but partly due to -or triggered by- external factors. Stellar distributions and peculiarities in HB morphology, such as gaps, and subluminous stars bluer than the canonical end of the HB -the so-called blue hook observed in the UV CMDs of the most massive GCs \citep[][]{1996ApJ...466..359D,1999AJ....118.1727P,1998ApJ...495..284W,Brown+2001,2007A&A...474..105B,2010ApJ...718.1332B}- may be important clues to study the HB star formation mechanisms. In particular, the formation of the extremely hot HB objects ($T_{eff} \geq$  32000 K) cannot be explained by the canonical stellar evolution and mass-loss theories. Interestingly, they were proposed to be the progeny of stars that, due to an unusually large mass loss, left the RGB before the helium flash and ignited helium later on during the white dwarf (WD) cooling sequence \citep{Castellani&Castellani93,1996ApJ...466..359D,Brown+2001}. Alternatively, they may be also the progeny of the helium enriched main sequence (MS) population, as proposed in the case of $\omega$-$Cen$ and NGC2808 \citep[e.g.][]{Lee2005}.}


In this paper we present a simple analytical model of the effects of stellar encounters on RGB mass-loss, and use it to predict the HB-mass distributions obtained from
samples of HB stars in the Galactic GCs NGC 5904, NGC 6093, NGC 6266, and NGC 6752. These clusters all show an Extended HB (EHB) morphology
\citep[][]{2007ApJ...661L..49L}, the presence of exotica likely originating from stellar interactions \citep[see e.g.][]{1997ApJ...482..870A, 2006AJ....131.2551B}, and an intermediate to old dynamical age, which may be an indication of dynamical effects playing a role in the RGB mass-loss mechanism \citep[][]{2013arXiv1305.1622P}.
In these environments our mass-loss law provides a remarkably good fit to HB-mass data, suggesting that tidal stripping through stellar encounters is probably the principal mechanism mediating the effect of dynamics on mass-loss. {A related issue is that of red-giant collisions and collisional depletion in dense clusters, as discussed e.g. by \cite{1996IAUS..174..355P, 1998ApJ...495..796S, 2004MNRAS.348..469A}}. We are {also} currently running Smoothed Particle Hydrodynamics (SPH) simulations of stellar collisions to produce a more detailed model that we will compare to a much larger database of HB mass distributions (Moraghan et al. in preparation).

{Our approach differs from previous studies \citep[e.g.][]{buonanno_horizontal_1997, recio-blanco_multivariate_2006, 2013arXiv1305.1622P} that mainly looked for correlations between a collisional parameter and a cluster's HB morphology indicator because it focuses on the shape of the mass-loss distribution of HB stars. The shape of the mass-loss distribution, as measured e.g. by the skewness parameter, is independent on the typical amount of mass-loss (as measured by the mean or median mass-loss) and on the dispersion in mass-loss (as measured by the standard deviation or other suitable estimators of dispersion). Thus we test a qualitatively different point with respect to previous studies, potentially complementing them \citep[see also the discussion in][who elaborates on this point from a slightly different perspective]{2013MmSAI..84...97D}.}

\section{Preliminary order of magnitude estimate}
\label{preli}
Before dealing with the full calculation, let us quickly estimate the fraction of RGB stars that undergo collisions resulting in significant mass-loss in a typical GC. If we call $r_{RGB}$ the radius of the RGB star, we can assume that encounters within distance $h r_{RGB}$ produce a significant mass-loss, where $h$ is a numerical factor. The cross-section for such an encounter will be $\sigma = \pi h^2 r_{RGB}^2$, neglecting both gravitational focusing effects and the radius of non-RGB stars. An RGB star orbiting in a GC core moves with average velocity $\nu$ during its RGB lifetime $t_{RGB}$. During this time it will have an encounter if a star is found within the cylindrical volume
\begin{equation}
\label{estimvolum}
V = \sigma \nu t_{RGB} = \pi h^2 r_{RGB}^2 \nu t_{RGB},
\end{equation}
so the probability of having an encounter is given by the number density $n$ times the volume $V$:
\begin{equation}
\label{estimprobab}
P = n \times \pi h^2 r_{RGB}^2 \nu t_{RGB}.
\end{equation}
Typical GC-core values are $\nu = 10$km/s, $n = 5 \times {10}^5/{pc}^3$, and for GC RGB stars\footnote{{The size of RGB stars varies over time during their evolution before reaching the RGB-tip value, so the following is an overestimate. A more accurate calculation would involve integrating the cross section for collisions over the RGB lifetime of a relevant stellar model.}} $t_{RGB} = 100$Myr and $r_{RGB} = 200$ $R_\odot$, so that
\begin{equation}
\label{estimprobabn}
P = h^2 \times 3 \times {10}^{-2}.
\end{equation}
Depending on the value of $h \approx 1 - 10$, the probability ranges from about some percent even up to $100\%$. It is difficult to constrain the expected probability in this rough approximation, because of the quadratic dependence on the unknown $h$. However, adopting $h = 2$, Eq.~\ref{estimprobabn} yields $12\%$, which is compatible with the typical fractions of EHB or Blue Hook stars found in GCs with an EHB morphology \citep{2011MNRAS.410..694D}. This is a hint that stellar collisions may play a role in the EHB formation mechanism, but a quantitative treatment is necessary if we want to obtain better constraints.

\section{The model}
\label{themodel}
We consider encounters between a single RGB star and a single cluster member (most likely a MS star). Initially we neglect gravitational focusing and approximate the orbits of both stars with straight lines, while in the next subsection we include the effects of gravitational focusing derived from a full solution of the two-body problem. It is clear that binary-single and binary-binary collisions (where the RGB or the scattering star, or both, were initially members of a binary) may play an important role in the actual mechanism of encounter-driven mass loss, because of the strong increase of the cross section of the interaction. However, EHB stars in clusters do not appear to be prevalently members of a binary system \citep[][]{2006A&A...451..499M, 2009A&A...498..737M, 2011A&A...528A.127M}. Therefore, considering the daunting complexity of the dynamics of such encounters \citep[see][]{1975AJ.....80..809H, 1993ApJS...85..347H} we decide not to address this issue in the present paper. To be sure, \cite{2012A&A...540A..16M} find a loose anticorrelation between the maximum HB temperature and MS binary fraction (as determined photometrically), so the issue is surely worthy of consideration and will be explored in detail in a follow-up paper.

\subsection{Negligible gravitational focusing approximation}
\label{negli}
We consider the impact parameter during a collision
\begin{equation}
\label{impact}
b = \left| \vec{x} - \left( \vec{x} \cdot \hat{v} \right) \hat{v} \right|
\end{equation}
where $\vec{x}$ and $\vec{v}$ are the scatterer star's position and velocity, and we assume that an encounter with impact parameter $b$ causes mass-loss when an RGB star overflows the associated Hill radius:
\begin{equation}
\label{Hill}
r_H = k \left( \frac{m_i}{M_{RGB}} \right)^{1/3} b
\end{equation}
where $k \approx 1$ is a numeric factor and $m_i/M_{RGB}$ is the mass ratio of the impactor to the RGB star. We approximate the density $\rho$ of the external layers of the RGB star as a power law in radius $r$, with the expression
\begin{equation}
\label{rholaw}
\rho(r) \propto r^{-11/3}.
\end{equation}
By integrating over the RGB star volume from $r_H$ to the surface, the overflowing mass we obtain is
\begin{equation}
\label{HilloverflowV}
\delta M = \int_{r_H}^{r_{RGB}} 4 \pi r^2 \rho(r) dr = A \left[ 1 - \left( \frac{b}{a} \right)^{2/3} \right]
\end{equation}
where $A$ represents the maximum amount of mass that can be lost in the event of a head-on collision, and
\begin{equation}
\label{HilloverflowVa}
a = \frac{r_{RGB}}{k} {\left( \frac{M_{RGB}}{m_i} \right)}^{1/3}
\end{equation}
is the maximum impact parameter for which mass-loss may take place. In the notation of Sect.~\ref{preli}, we can write
\begin{equation}
\label{withh}
h = \frac{1}{k} {\left( \frac{M_{RGB}}{m_i} \right)}^{1/3} \gtrapprox 1
\end{equation}
Since Eq.~\ref{rholaw} holds only in the external layers, in general $A \neq M$, so we allow for it to be a free parameter. We also introduce a zero-point correction parameter $B$, possibly related to systematic uncertainties in the determination of the RGB star mass before the encounter. So Eq.~\ref{HilloverflowV} becomes
\begin{equation}
\label{HilloverflowVpar}
\delta M = A \left[ 1 - \left( \frac{b}{a} \right)^{2/3} \right] + B
\end{equation}
We model the phase-space distribution of stars in a GC as isotropic and constant-density. {We will consider the impact parameter distribution $f(b)$ within a sphere of radius $a$, i.e. under the condition that $b < a$ and the assumption that stars with $b < a$ uniformly populate the $r < a$ sphere. The resulting impact parameter distribution is expressed as:}
\begin{equation}
\label{fb}
f(b) = \frac{3 b \sqrt{1 - \frac{b^2}{a^2}}}{a^2}
\end{equation}
{This is a zeroth-order approximation, given that the number density of stars in the cluster actually varies by orders of magnitude when moving along an eccentric stellar orbit that dips into the GC core and reaches to the outskirts over a timescale that is potentially shorter than RGB lifetime, but the assumption that underlies our result is increasingly accurate for decreasing $b$ (i.e. increasing mass-loss), so we expect environmental factors to affect the low-mass-loss part of our distribution more than its high-mass-loss tail.
Notice that at small $b \ll a$, $f(b)$ is simply linear. Indeed, the parameter $a$ can be seen as a regularizer of the distribution for large $b$ and at small $b$ one 
recovers $f(b) db \propto 2\pi b db$. It is useful to consider also the cumulative distribution function $F(b)$, defined as the probability of having an impact parameter smaller than $b$. We obtain
\begin{equation}
\label{Fb}
F(b) = 1-\left(1-\frac{b^2}{a^2}\right)^{3/2}
\end{equation}
and then for $b\ll a$, the expected quadratic behavior is recovered $F(b) \propto b^2/a^2$.
}

\begin{figure}
\includegraphics[width=0.8\columnwidth]{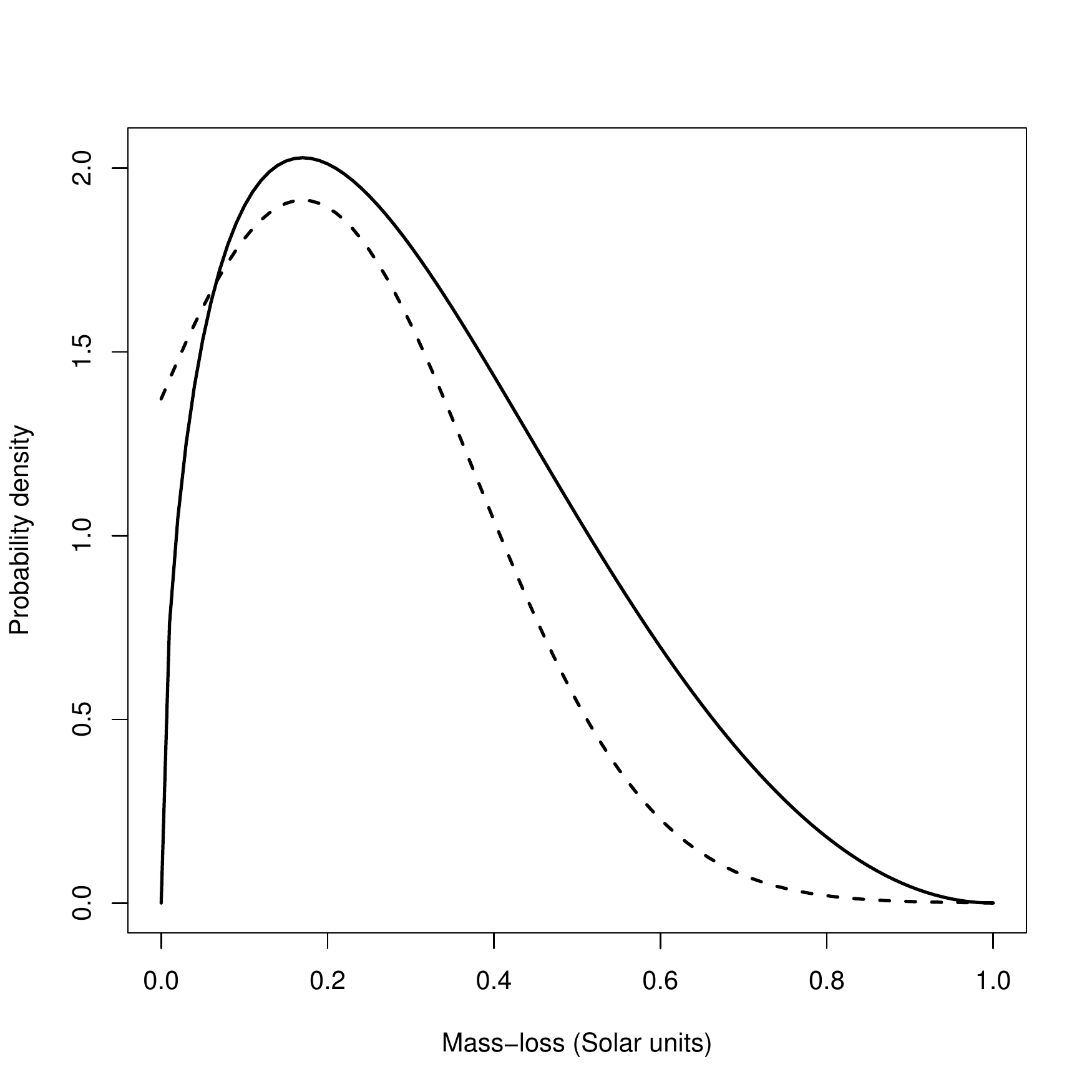}
\caption{Solid line is the distribution of Eq.~\ref{Hilloverflowbetter} for $A = 1$ and $B = 0$. Dashed line is a Gaussian with same FWHM and mode. Notice how our model shows an heavy tail towards high mass-loss compared to the Gaussian.\label{dispo}}
\end{figure}

Finally, combining Eq.~\ref{fb} with Eq.~\ref{HilloverflowVpar} we obtain the probability distribution of $\delta M$ as\footnote{Given the theoretical distribution of impact parameters in Eq.~\ref{fb} and the relation to mass-loss assumed in Eq.~\ref{HilloverflowVpar} there are several ways to obtain Eq.~\ref{Hilloverflowbetter}. For example, we can notice that  Eq.~\ref{HilloverflowVpar} is monotonic so that impact parameters larger than a given value $b$ correspond to mass-loss smaller than $\delta M (b)$. In this way, the cumulative distribution of $\delta M$ can be obtained by applying Eq.~\ref{HilloverflowVpar} to the cumulative distribution of $1 - b/a$. Afterwards,  Eq.~\ref{Hilloverflowbetter} can be obtained by derivation with respect to $\delta M$. This procedure is general and applies to any distribution of impact parameters and to any mass-loss law that is monotonic in the impact parameter.}
\begin{equation}
\label{Hilloverflowbetter}
g(\delta M) = \frac{9}{2 A} {\left(1 - \frac{\delta M - B}{A}\right)}^2 \sqrt{1 - {\left[ 1 - \left( \frac{\delta M - B}{A} \right) \right]}^3}
\end{equation}
The distribution lacks an adjustable shape parameter: changes in the values of the parameters $A$ and $B$ correspond to a linear rescaling\footnote{For the normalization to be preserved, however, $A$ has to also appear in the denominator of Eq.~\ref{Hilloverflowbetter}.} and a rigid translation respectively, and they do not affect the shape of the curve which is skewed towards high values of mass-loss. We show the shape of the distribution for $A = 1$ and $B = 0$ in Fig.~\ref{dispo}, where a Gaussian distribution with the same FWHM and mode (maximum) is also plotted for comparison. Notice how the tail of the Gaussian distribution drops considerably faster for increasing values of mass-loss with respect to our model.

Note that Eq.~\ref{Hilloverflowbetter} describes the distribution of mass-loss only for stars that underwent at least a collision, i.e. an encounter with $b/a < 1$. We also neglect multiple collisions so that at least a collision means exactly one collision\footnote{This is inconsistent when $P \to 1$ in Eq.~\ref{estimprobabn} because the probability of double (independent) collisions scales as $P^2$. However we assume that the cross section is considerably reduced after the first collision, as is the case when most of the envelope mass is stripped.}. The probability of having a collision does not enter Eq.~\ref{Hilloverflowbetter}, and it is impossible to use it to estimate the maximum impact parameter $a$ resulting in mass-loss. The probability $P$ of colliding is irrelevant to the shape of the distribution of mass-loss for those stars that actually collided, and vice-versa. The full mass-loss distribution, including stars that did not collide, is
\begin{equation}
\label{letsuseadeltafunction}
g_{\mathit{all}}(\delta M) = P g(\delta M) + (1 - P) \delta_D{(\delta M)}
\end{equation}
where $\delta_D$ is the Dirac delta function centered in $0$. In a real CMD, however, photometric errors, differential reddening and mass-loss mechanisms not related to encounters would broaden the delta function. To model this, we would need to introduce further parameters, limiting the predictivity of our model and leading to a more complex analysis that is beyond the scope of this paper. {In the following, however, we will assume $P = 1$ and neglect the Dirac delta term, even though this is justified only in extremely high-density environments where, for a given $h \approx 2$ the probability of collision $P$ can be inflated by a larger numeric coefficient with respect to that in Eq.~\ref{estimprobabn}, that refers to average cluster values}.

\subsection{Full model with gravitational focusing}
\begin{figure}
\includegraphics[angle=270, width=0.8\columnwidth]{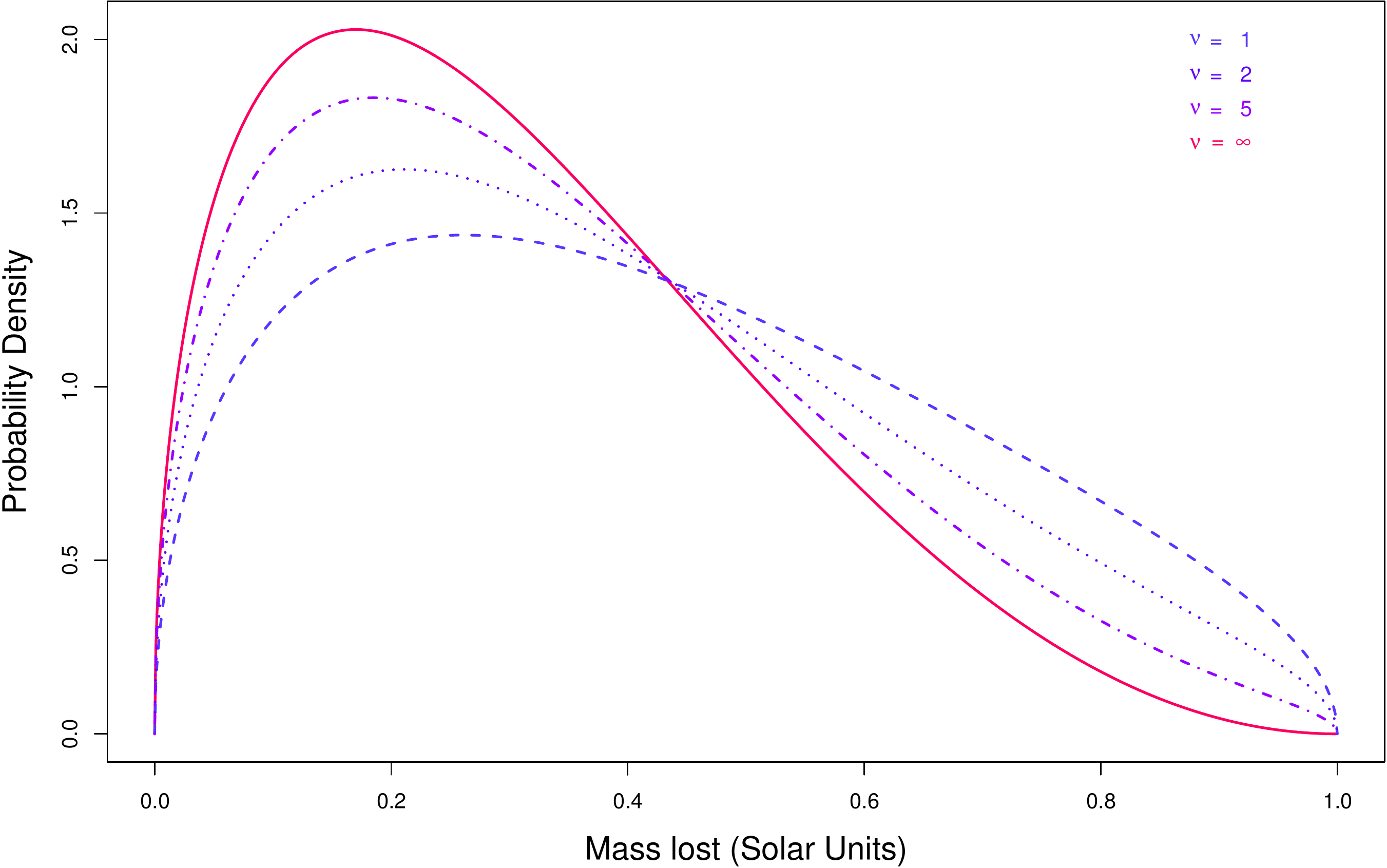}
\caption{Red solid line is the distribution of Eq.~\ref{Hilloverflowbetter}
for $A = 1$ and $B = 0$, {corresponding to the limit of
Eq.~\ref{sticatshee} for which gravitational focusing is negligible,
i.e. $\nu \to \infty$. Dot-dashed, dotted and dashed lines in
progressively darker shades of purple correspond to lowering values
of the dimensionless parameter $\nu$ in Eq.~\ref{sticatshee} (see
Eq.~\ref{definu} for the definition of $\nu$), i.e.  $\nu=5,2,1$,
respectively. The shape of the distribution approaches the $\nu \to
\infty$ limit already for $\nu = 1$, which is a typical value in GCs
if we assume a value of the parameter $a$ in Eq.~\ref{definu} of
order several times the RGB radius.}\label{gra}}
\end{figure}

{We now add gravitational focusing to the simple model outlined above. The periastron during an encounter is a function of the impact parameter and relative velocity:
\begin{equation}
\label{rofb}
 r(b,v) = \frac{2 G m \left(\sqrt{\frac{b^2 v^4}{4 G^2 m^2}+1}-1\right)}{v^2} \;.
\end{equation}
{where $m = m_i + M_{RGB}$ is the sum of the two star's masses}, and consistency with the previous section is insured given that $r(b,v) \stackrel{v\to \infty}{\longrightarrow} b$. {The distribution of the impact parameter is, as before, given by Eq.~\ref{fb},
but now we assume that the relevant parameter for mass loss is the periastron. Therefore the mass lost is now a function of the periastron, i.e.
\begin{equation}
\label{masslostR}
\delta M (r) =  A \left[1-\left(\frac{r}{r_a}\right)^{2/3}\right]+B
\end{equation}
where the maximum radius at which we have mass lost is $ r_a \equiv r(a) $. The distribution of $r$ can be obtained by combining \ref{rofb} and \ref{fb}
\begin{equation}
 P(r) = \frac{3 (2 \nu  \rho +1) \sqrt{\nu -\nu  \rho ^2-\rho }}{2 a \nu^{3/2}}
\end{equation}
while its cumulative distribution, defined as the probability of having a periastron smaller than $r$, takes the form
\begin{equation}
 \label{cumr}
 C(r) = 1-\left(1-\frac{\rho  (\nu  \rho +1)}{\nu }\right)^{3/2}
\end{equation}
where we set $\rho = r/a$ and
\begin{equation}
\label{definu}
\nu = \frac{a v^2}{4 G m}.
\end{equation}
The {dimensionless} quantity $\nu$ is key to the following discussion, in that it parameterizes the strength of gravitational focusing. It {is proportional to} the ratio between the typical velocity of stars $v$ and a scale escape velocity from the RGB star (the scale being set by $a$, {the maximum impact parameter for which mass-loss occurs, which is a free parameter in our model}), squared. {The condition $\nu > 1/2$ holds for hyperbolic orbits. Typical values in GC cores are of order $1$ for $a$ of order several times the RGB radius.} Clearly $P(r) \to f(r)$ for $\nu \gg 1$, and gravitational focusing becomes negligible in this limit recovering our previous results. 
On the other hand, if the gravitational focusing is not negligible and $\nu$ is finite, the behavior at small $r$ is strongly modified. In fact, in the relevant region at small periastron, i.e. $\rho \ll \nu$, the cumulative distribution of the periastron becomes linear  $C(r) \simeq \frac{3\rho}{2\nu}$. This 
should be compared with the quadratic behavior at small $b$ that we had in Eq.~\ref{Fb}.}

The general distribution of the mass lost, according to Eq.~\ref{masslostR}, is then given by
\begin{equation}
\label{sticatshee}
P(\delta M) =  \frac{9 \left(u^{3/2} (z-1)+1\right) \sqrt{u \left(z+1-2 u^{3/2}-u^3
   (z-1)\right)}}{2 A (z+1)^{3/2}}	
\end{equation}
where $u = \frac{A + B - M}{A}$ and $z = \sqrt{1 + 4 \nu^2}$.
In Fig.~\ref{gra} we compare the shape of the distribution predicted by Eq.~\ref{sticatshee} for various values of $\nu$ with the limit $\nu \to \infty$.
{The main effect of gravitational focusing is, predictably, to thicken the high-mass-loss tail and to make the distribution somewhat more symmetric, lowering its skewness. The physical interpretation of this occurrence is that gravitational focusing increases the proportion of encounters that remove a large amount of mass in the distribution. However, it is easy to notice that even for low values of $\nu$ (of order unity to several) the skewed shape of the distribution that is distinctive of our result is still readily visible.}
In the following sections, for simplicity, we limit our quantitative comparison to the data to the $\nu \to \infty$ limit. However, we will see that the cluster most affected by gravitational focusing effects (NGC 5904; see Table \ref{tab:dynGCs}) deviates the most from the expectations of Eq.~\ref{Hilloverflowbetter}, and precisely in the direction of increased symmetry in the distribution. Meanwhile, NGC 6266, the least affected, shows the best agreement with Eq.~\ref{Hilloverflowbetter}.
}

\section{Testing the model on globular clusters}
In order to test our mass-loss law in detail we consider four GCs, namely NGC 5904, NGC 6093, NGC 6266, and NGC 6752. {See Table \ref{tab:GCs} and Table \ref{tab:dynGCs} for a summary of the sample properties}. It has become clear in recent years that GCs are not simple stellar populations, but may host chemically distinct subpopulations, typically with an enhanced helium abundance. These helium-rich populations can make up a substantial fraction of all cluster stars, and can play a role in shaping the HB morphology as well \citep[e.g.][]{2010A&A...517A..81G}.
However, it should be clear that in this section we are exploring the capability of our model to reproduce the HB morphology without including other mechanisms able to modify the distribution of the stars in the HB.

The clusters in our sample are chosen in order to have well or moderately extended blue HB tails. NGC 5904 and NGC 6266 possess a similar metal content, but the latter presents a HB more extended toward high temperatures. In NGC 5904 the $c_{U, B, I}$ distribution of RGB stars confirms the multimodality of the red giant branch \citep{2013MNRAS.431.2126M}.

NGC 6093 (M80) is one of the densest GCs in the Milky Way. Apart from GC abundance anomalies, there is no photometric evidence of multiple stellar populations to date, as recently shown by \citet{2013MNRAS.431.2126M} who do not find multiple RGB sequences using the $c_{U,B,I}$ color.

NGC 6752 is a moderately metal poor cluster with [Fe/H]$\sim - 1.55$ \citep{2005A&A...438..875Y, 2010A&A...516A..55C}, with large star-to-star variations in O, N, Na, Mg, and Al. The Na-O and Mg-Al anticorrelations have been observed by \citet{2005A&A...438..875Y,2012ApJ...750L..14C}. Recently, in this cluster three main populations of stars have been photometrically identified by \citet{2013MNRAS.431.2126M}.

For each cluster, the starting values of reddening and distance are taken from the Harris's catalog\footnote{\hbox{\url{http://www.physics.mcmaster.ca/Globular.html}}, v. 2010} \citep{1996AJ....112.1487H}, except for NGC 6093, for which we assumed values from \citet{2005A&A...432..851R}. Note here, that we neglect a possible broadening of the CMDs due to differential and non-uniform reddening across the cluster. For instance NGC 6266 is located in the direction of the galactic center, and its extinction across its visible surface is nonuniform causing the features in its color-magnitude diagram to be broadened \citep{2011AJ....141..146A}.

Observations were taken from the HST-WFPC2 database of \citet{2002A&A...391..945P}\footnote{\hbox{\url{http://www.astro.unipd.it/globulars}}}, except for NGC 6752 from the HST-ACS data by \citet{2007AJ....133.1658S}\footnote{\hbox{\url{http://www.astro.ufl.edu/~ata/public_hstgc/databases.html}}}. We derive the best fit for each cluster by assuming [Fe/H] from the metallicity scale by \citet{2009A&A...508..695C}, see Table \ref{tab:GCs}. We produce synthetic CMDs from the main-sequence phase up to the asymptotic branch using the stellar population synthesis code SPoT (Stellar POpulation Tools, see for details and ingredients \citealt{2005AJ....130.2625R,2009ApJ...700.1247R}). The metallicity values we adopt are $Z = 0.001$ for NGC 5904 and NGC 6266, and $Z = 0.0006$ for NGC 6093 and NGC 6752. As for ages, the best fit is found at an age of 13 Gyr in the case of NGC 6093, and 12 Gyr for NGC 5904 and NGC 6266, all in agreement with the relative age scale of \citet{2009ApJ...694.1498M}. An age range spanning from 11 Gyr up to 14 Gyr is also taken into account in order to explore the effect of age on the mass at the tip of the RGB. Our model in any case is sensitive to the RGB-tip mass only through the shift-parameter $B$, while the overall shape of the mass-loss distribution function is the same for different values of the mass at the tip of the RGB.

\begin{figure}
\includegraphics[width=0.8\columnwidth]{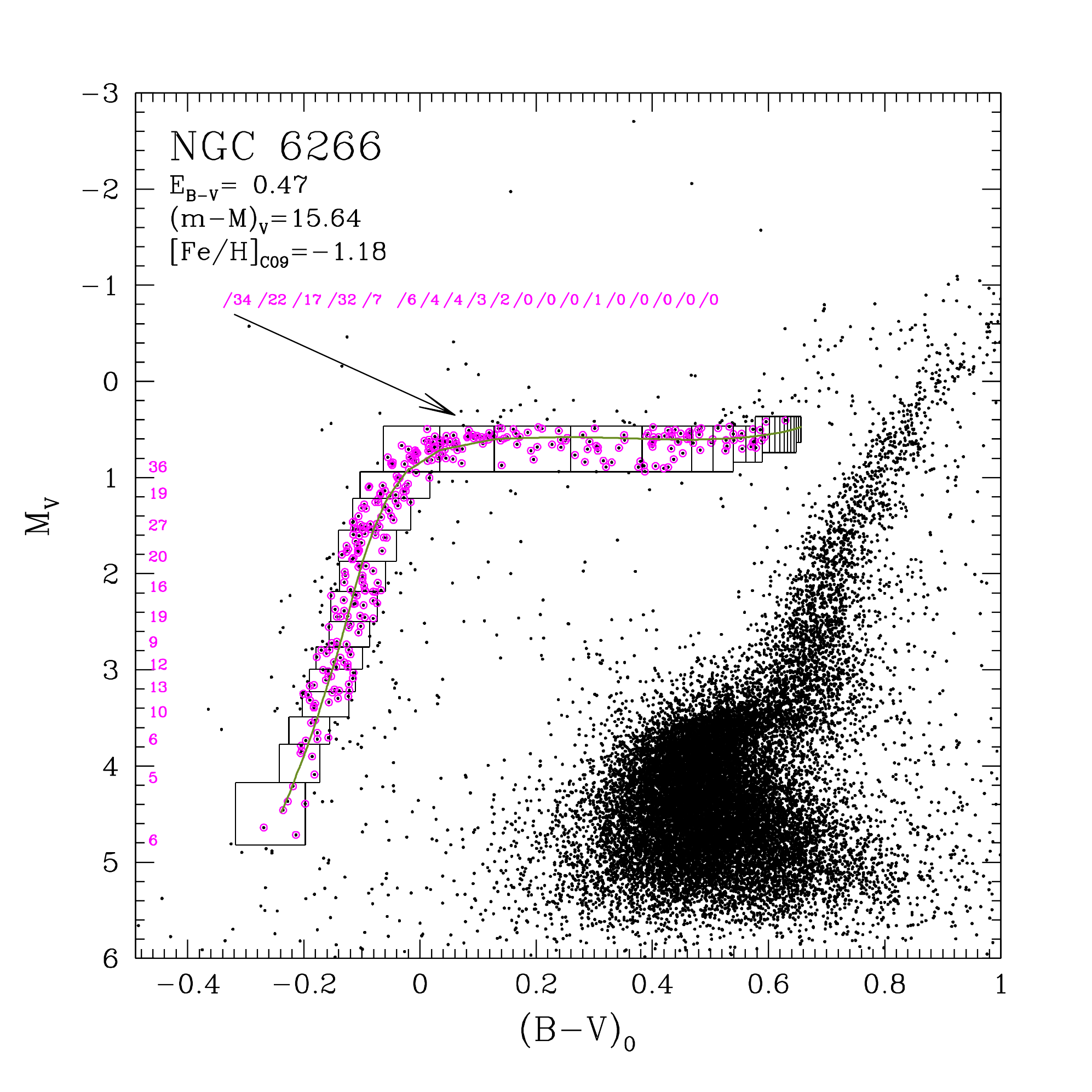}
\caption{Observed color-magnitude diagram of NGC\,6266. The ZAHB location (dark green line) for classical HB evolutionary models of $Z = 0.001$ is shown. Boxes correspond to mass intervals of $\delta M = 0.01 M_\odot$ that we used to derive the histogram of mass distribution along the observed HB. \label{fig6266}}
\end{figure}

The ZAHB from the best-fit model is used to derive the mass distribution along the HB, by counting stars in $0.01$ solar-mass bins along it, as illustrated on the CMD of NGC 6266 in Fig.~\ref{fig6266}. 
The entire synthetic CMD and mass values are from canonical theoretical stellar evolution models \citep[][and references therein]{2006ApJ...642..797P}. This choice is justified by the fact we are looking at a pure dynamical effect of mass loss on the HB mass distribution. We plan to include the effects of Helium-enhancement, age, chemistry, and possibly further parameters in a forthcoming paper, alongside with refinements of our model.

\begin{deluxetable}{llllll}
\centering
\tablecaption{Properties of our comparison clusters.\label{tab:GCs}}
\tablewidth{0pt}
\tabletypesize{\footnotesize}
\setlength{\tabcolsep}{0.07in}
\tablehead{ \colhead{Name} & \colhead{ M$_V$ } & \colhead{$c$} & \colhead{E$_{B-V}$} & \colhead{$(m-M)_V$} & \colhead{$[Fe/H]$} }
\startdata
NGC 5904 (M5)  & -8.81 & 1.73  & 0.02 & 14.46 & $-1.33\pm 0.02$\\
NGC 6093 (M80) & -8.23 & 1.68  & 0.03 & 15.75 & $-1.75\pm 0.08$ \\
NGC 6266 (M62) & -9.18 & 1.71c & 0.03 & 15.64 & $-1.18\pm 0.07$ \\
NGC 6752       & -7.73 & 2.50c & 0.06 & 13.15 & $-1.55\pm 0.01$
\enddata
\tablecomments{Columns: cluster identification; total cluster magnitude in V, central concentration, reddening, and distance modulus from Harris's catalog; [Fe/H] from  \citet{2009A&A...508..695C}.}
\end{deluxetable}

\begin{deluxetable}{llllll}
\centering
\tablecaption{Relevant dynamical properties and maximum HB temperature of our comparison clusters.\label{tab:dynGCs}}
\tablewidth{0pt}
\tabletypesize{\footnotesize}
\setlength{\tabcolsep}{0.07in}
\tablehead{ \colhead{Name} & \colhead{ $\sigma_K$ } & \colhead{$\Gamma_{coll}$} & \colhead{$\nu$} & \colhead{$f_b$} & \colhead{$\log{T^{max}_{HB}}$}}
\startdata
NGC 5904 (M5)   & $9.01$ & $-14.30$  & $1.0$&  $0.022\pm0.006$ &$4.176$\\
NGC 6093 (M80) & $10.37$ & $-13.68$ & $1.3$&  $0.012\pm0.006$ &$4.477$\\
NGC 6266 (M62) & $16.57$ & $-14.22$ & $3.4$&  NA &$4.477$\\
NGC 6752           & NA &  NA & NA & $0.010\pm0.006$ & NA
\enddata
\tablecomments{Columns: cluster identification; velocity dispersion from King model fit from \cite{mclaughlin_resolved_2005}, collisional parameter from Table 1 of \cite{recio-blanco_multivariate_2006}; gravitational focusing parameter $\nu$ as defined in Eq.~\ref{definu}, normalized to NGC 5904; total MS binary fraction ($f_{bin}^{TOT}$ in WFC field) from Table 2 of \cite{2012A&A...540A..16M}; and maximum HB temperature.}
\end{deluxetable}


\subsection{Results}
{We obtained a simple physical model of collision-mediated RGB mass-loss. Its main prediction is the asymmetry of the mass-loss probability distribution function, due to the fundamental fact that distant ecounters, that result in lower mass-loss, are more frequent than near encounters, that produce high mass-loss. This is a feature that we expect to be shared by all collision-based models, whatever the details of how the mass-stripping interaction is modeled. Instead, a Gaussian mass-loss distribution, is symmetric by construction. In our four clusters we obtained the mass-loss distribution by subtracting the HB mass to the mass at the tip of the RGB as obtained self-consistently by the best-fit to the CMD. We then tested the Gaussianity of the mass-loss distribution thus obtained by using a Shapiro-Wilk Gaussianity test \citep[][]{Royston1982} and found that at different confidence levels (see the p-values listed in Col.~5 of Table \ref{addita}) it can be rejected.}
 
{To measure the amount of symmetry observed in the empirical dataset indepenently from the Gaussianity hypothesis, we introduce a robust skewness parameter based on the sample median ($q_{50\%}$) and $25\%$ and $75\%$ quantiles ($q_{25\%}$ and $q_{75\%}$), defined as
\begin{equation}
\label{qsk}
s = \frac{1}{2} - \frac{q_{50\%} - q_{25\%}}{q_{75\%} - q_{25\%}}
\end{equation}
obtaining the values listed in Col.~6 of Table \ref{addita}. The uncertainty on $s$ is obtained by bootstrap resampling. A value of $s = 0$, as would result from a symmetric distribution, is compatible with the data in only one case (NGC 5904)\footnote{This does not contradict the Shapiro-Wilk test result. Apparently NGC 5094 fails the S-W test because of either asymmetry in the tails of the distribution further from the median than the first and third quartile, which are not considered by our robust skewness parameter, or due to non-zero kurtosis or issues with higher momenta.} while the other clusters have an asymmetric distribution, as expected {if our model applies}. The reason why NGC 5094 has a more symmetric distribution than the other clusters, at least in the central region probed by our robust indicator of skewness, may be related to the effect of gravitational focusing, but a more in-depth analysis would be required to settle the case.}

In Fig.~\ref{fithist} we show the best-fit curve from our model superimposed onto the NGC 6266 mass-loss data, {together with the Gaussian best fit}. The best fit curve (with $A =  0.222 \pm 0.007$, $B = 0.158 \pm 0.002$; we describe the fitting procedure in Appendix~\ref{appe}) reproduces the shape of the observed mass-loss datapoints to a high degree of accuracy. {This is evidence that in NGC 6266, stellar encounters may be able to act as the main ingredient in determining RGB mass-loss}. However note that in Fig.~\ref{fithist} the bins corresponding to low mass-loss, below the cutoff predicted by our model, contain a small number of stars anyway. Since our model predicts strictly $0$ stars there, this would result in a formal rejection of the model in statistical tests such as the Pearson's $\chi^2$ test (see below). To justify this discrepancy, stellar evolution off the ZAHB may be invoked. Evolved HB stars, i.e. stars that no longer lie on the ZAHB, may appear as higher-mass ZAHB stars, because evolutionary tracks run almost parallel to the ZAHB from the blue to the red side of the CMD as stellar age increases, resulting in a contamination of the high-mass bins with lower-mass, evolved stars.

In NGC 6093, NGC 5904, and NGC 6752 the agreement between the mass-loss data and our model is somewhat less striking, but the main features (the positive skewness, left cutoff, and right long tail) are well matched, with the exception, as discussed above of NGC 5904. We point out that this cluster has a relatively low central density with respect to the others in the sample, more than an order of magnitude less than NGC 6266, according to \cite{1996AJ....112.1487H}, {and, as anticipated above, gravitational focusing is about three times more important in NGC 5904 compared to NGC 6266 (see $\nu$ in Table \ref{tab:dynGCs})}. On the other hand, the collision parameter $\Gamma_{coll}$ is similar for the two clusters despite the differences in the shape of the mass-loss distribution. This is not surprising, given that the collision parameter determines the probability of a star colliding, but does not influence the details of what happens during the collision. It is precisely those details that dictate the shape of the mass-loss distribution function. Figure \ref{theothers} shows our best-fit models superimposed to the histograms of mass-loss for the HB of these GCs,  {together with the Gaussian best fit}. We list in Table \ref{addita} the relevant fit parameters and uncertainties (Col~3, Col~4), and HB-star sample size (Col.~2).

As a further check that our results are not unduly influenced by off-ZAHB evolution we also compared our model to a Gaussian only in the high mass-loss tail of the distribution, i.e. $\delta M > 0.16$. We choose this threshold because it is larger than the low-mass cutoff in all the clusters we considered. It is also interesting to test the goodness of fit of our model in the high-mass-loss tail of the distribution (corresponding to low impact parameters in Eq.~\ref{fb}) because it is the least affected by environmental factors as discussed in Sect.~\ref{negli}. {For each GC in our sample we obtained mean and standard deviation estimates from the whole dataset, i.e. before applying any cutoff to the mass-loss distribution. We then plugged these values into a Gaussian function, and we compared it to our model, whose parameters have been similarly obtained by fitting the whole dataset. The comparison was carried out only on the datapoints above the $\delta M > 0.16$ cutoff through a {Pearson's} $\chi^2$ test, i.e. by computing
\begin{equation}
\label{chisqu}
\chi^2 = \Sigma_i \left[ \frac{\left( {N_{\mathit{observed}, i} - N_{\mathit{predicted}, i}} \right)^2}{N_{\mathit{predicted}, i}} \right]
\end{equation}
over the bins above the $\delta M > 0.16$ cutoff\footnote{{If we compared the models over the full range of the observations, our model would have $\chi^2 = \infty$ due to its prediction of strictly $0$ points in the left tail.}}. In Table \ref{addita2} we report the values of $\chi^2$ obtained for the two distributions, i.e. either using our distribution or the Gaussian to predict the number of observations in each bin. Note that both distributions have two free parameters and are compared on the same set of datapoints and bins, consequently sharing the same number of degrees of freedom, which results in values of $\chi^2$ that can be compared directly}. It can be seen that in the case of NGC 5904 the Gaussian model {has a lower $\chi^2$, thus outperforming our distribution function, while in the other cases we have either similar values (in NGC 6266) or a marked superiority of our model (i.e. a lower value of $\chi^2$) in NGC 6093 and 6752. In Table \ref{addita2} we also list the error bars on $A$ and $B$ obtained by bootstrap resampling (with replacement) the observed mass-loss values for each cluster, and repeating the fitting procedure $10000$ times. The associated standard deviation of the sample of fitting parameters thus obtained is compatible with the error bars on fit coefficients obtained through the standard linear regression procedure as listed in Table \ref{addita}.}

\begin{deluxetable}{cccccccc}
\centering
\tablecaption{Summary of our results on our comparison clusters.\label{addita}}
\tablewidth{0pt}
\tabletypesize{\footnotesize}
\setlength{\tabcolsep}{0.07in}
\tablehead{ \colhead{Name} & \colhead{N} & \colhead{$A$} & \colhead{$B$} & \colhead{$S-W$} & \colhead{skew} }
\startdata
NGC 5904 & $284$ & $0.16 \pm 0.02$ & $0.158 \pm 0.005$& $7.3 \times {10}^{-11}$ & $0.00$\\
NGC 6093 & $328$ & $0.24 \pm 0.03$ & $0.143 \pm 0.009$& $3.4 \times {10}^{-14}$ & $0.17$\\
NGC 6266 & $309$ & $0.222 \pm 0.007$ & $0.158 \pm 0.002$ & $ 2.8 \times {10}^{-4}$ & $0.17$\\
NGC 6752 & $120$ & $0.29 \pm 0.02$ & $0.166 \pm 0.006$& $7.9 \times {10}^{-7}$ & $0.17$
\enddata
\tablecomments{Fit parameters with associated uncertainties ($A$, $\Delta A$, $B$, $\Delta B$), Shapiro-Wilk test p-value (S-W p), and quantile-based skewness parameter (skew). The $B$ coefficients are compatible within $2$-$\sigma$ error bars, while the $A$ coefficients are not.}
\end{deluxetable}

\begin{deluxetable}{cccccccccc}
\centering
\tablecaption{Bootstrapping estimates of errors on our model's coefficients and $\chi^2$ test results for the high-mass-loss tail.\label{addita2}}
\tablewidth{0pt}
\tabletypesize{\footnotesize}
\setlength{\tabcolsep}{0.07in}
\tablehead{ \colhead{Name}  & \colhead{$\Delta A$} & \colhead{$\Delta B$} & Mean & S. d. & \colhead{$\chi^2$} & \colhead{$\chi^2$ Gaussian} & d. f.}
\startdata
NGC 5904 & $0.01$ & $0.004$& $0.21$ & $0.04$ & $81.2$ & $53.2$ & $3$\\
NGC 6093 & $0.02$ & $0.003$& $0.22$ & $0.05$ &$37.7$ & $95.1$ & $3$\\
NGC 6266 & $0.01$ & $0.004$ & $0.22$ & $0.05$ &$18.3$ & $14.2$ & $3$\\
NGC 6752 & $0.02$ & $0.006$& $0.25$ & $0.05$ &$20.5 $ & $37.9$ &$3$
\enddata
\tablecomments{Fit parameter uncertainties as estimated from a bootstrap resampling of the mass-loss sample and refitting ($\Delta A$, $\Delta B$), {outcome of the Pearson's $\chi^2$ test} for our model and for a Gaussian (see text), {and relevant degrees of freedom. We also include the standard deviation and mean of the samples, that we used as parameters for the Gaussian}.}
\end{deluxetable}

Interestingly, Figures \ref{fithist}-\ref{theothers} show that RG stars need to collisionally lose a mass as high as $\sim 0.3-0.35 M_\sun$ in order to account for the presence of hot HB stars. These mass-loss values are fully compatible with the minimum total mass required for a star to be on the hot part of the HB. Since the early '70s it was clear that a star ignites He with a violent flash at the RGB tip if the survived H-rich envelope mass is a few hundredths of a solar mass over the He core \citep[e.g.][in the case of mass exchange in a binary system]{Giannone70}.
On the contrary, if stars undergo very high mass loss and the H-rich envelope is reduced to a few thousandths of a solar mass, they leave the RGB and ignite He with a late flash while descending the WD cooling curve \citep[e.g.][]{Castellani&Castellani93, 1996ApJ...466..359D,Brown+2001}. Moreover, during this phase hot HB stars can experience a flash-induced mixing inside the He core able to reach the H-rich envelope \citep[see e.g.][]{ 2003ApJ...582L..43C} with observational consequences on their surface abundances \citep{moni2012}.  All these previous investigations assumed the Reimers' formalism to describe the mass-loss process by changing the Reimers' parameter ($\eta_R$) up to 1 or more. This can be defined as a "slow" mass-loss mechanism since the mass loss progressively increases as stellar luminosity increases and gravity decreases. In order to reproduce the collision picture, we computed new evolutionary models with the stellar evolution code ATON \citep{2008Ap&SS.316...93V} by applying an instantaneous (episodic) mass stripping ($\delta M_{coll}$) to a star with an initial mass of $M=0.8 M_\sun$, $Z = 0.0006$ and $Y = 0.24$. In the models, we assumed that the collision occurs when the helium core mass is $M_{core} \simeq 0.45 M_\sun$ and lasts nearly 100 yr, and different values of $\delta M_{coll}$. In Fig. \ref{fig:evolution} three representative cases are shown. After the collision event, the star reacts to the instantaneous mass loss and attempts to  re-arrange the structural quantities of its envelope according to the instantaneous total mass. When $\Delta M_{coll} \sim 0.36 M_\sun$ (bottom panel) the star does not succeed in igniting helium, due to the very low residual envelope mass. If the stripped mass is slightly lower, e.g. $\delta M_{coll} \sim 0.26$, the star rapidly crosses the Hertzsprung-Russell (HR) diagram towards high effective temperatures ($\log T_{eff}\sim 4.9$) before experiencing a late He flash. Nevertheless, the survived H-rich layer is so small that it is rapidly exhausted by the shell above the He-rich core, the He burning switches off and the star becomes a WD. For a collisional mass-loss lower than $\sim 0.2 M_\sun$, the star experiences the He flash at high or low effective temperatures proceeding to a "normal" He-burning evolutionary sequence. For the aforementioned ages and values of metal content, the RGB-tip star mass is of the order of 0.80 $M_\sun$ and the He-core mass at the He-flash is $\sim 0.5 M_\sun$. Therefore, if after the collision the survived envelope mass is $\gsim 0.05 M_\sun$, the subsequent evolution does not show substantial changes after a short time needed to rearrange all its structural quantities.  A more complete study of this matter is beyond the purpose of this paper, but we plan to deepen this issue in a forthcoming paper.

\section{Conclusions and future prospects}
We presented a simple dynamical model for enhanced mass-loss in the RGB phase through tidal stripping in stellar encounters. The model predicts a distinctively {asymmetric distribution of mass-loss, with a tail towards high values of mass-loss}. The resulting distribution function contains only two freely adjustable parameters that regulate the overall amount and the spread in mass-loss, while the distribution shape is fixed.
We checked our model against the distribution of HB mass in $4$ Galactic EHB GCs from HST CMDs. Given an initial RGB-tip mass, this results in a distribution of mass-loss in the RGB phase that we compared to our model. {The obtained mass-loss distribution is skewed, except for NGC 5904, and a Shapiro-Wilk test suggests that the distribution is not Gaussian}. {Our simple dynamical model of mass-loss reproduces the shape of the mass-loss distribution with satisfactory to excellent accuracy, showing that stellar encounters may act as the main ingredient in determining RGB mass-loss}.

Both photometric datasets we used refer to GGC core regions. In a more extended comparison between our model and observations, a difference in the stellar number ratio between EHB and red HB stars in the less collisional GC outskirts and its core might be possible, even tough this will depend on star orbits and the cluster size. For a typical GC mass ($10^5 M_\sun$) and half-mass radius (3 pc), the time scale on which orbits in the cluster mix (i.e. the crossing time) is of the order of $\sim 3.5 \times 10^5$ yr, which is very much smaller than the GGC ages. Results from radial distribution studies and number ratios in the center and outskirts of GCs are still an open issue. A study on NGC2808 by \citet{iannicola2009} has revealed that the relative fractions of cool, hot, and extreme HB stars do not change radically when moving from the center to the outskirts of that cluster. The authors argued against the presence of strong radial differentiation among any stellar subpopulations having distinctly different helium abundances. Interestingly, the ratio of HB to RGB stars brighter than the ZAHB luminosity level steadily increases when moving from the innermost to the outermost cluster regions, a possible indication of a deficiency of bright RGB stars in the outskirts of the cluster. \citet{nataf2011} showed that the HB of 47 Tuc becomes both fainter and redder for sightlines farther from the cluster center. They evidenced that a gradient is found to be significant in the cluster outskirts, while closer in, the data are more compatible with a uniform mixing. If we speculate on this feature in terms of mass-loss, heavy HB stars, i.e. those whose RG progenitors have lost a small amount of mass, are preferably located out of the center. Of course, we do not intend to correlate this observational evidence with our collision scenario, since many factors can contribute to the picture, such as differential reddening, multiple generation of stars, and different helium abundances, but rather as a hint for further study such as with photometric data obtained by new wide-field imagers {\citep[e.g. OmegaCam on the VST Telescope;][]{2011Msngr.146....2C}.}

In a previous paper we showed that, while HB morphology is mainly determined by metallicity, age, and Helium enhancement, dynamical effects most likely are present, especially in high-density clusters suspected to be undergoing core-collapse \citep[][]{2013arXiv1305.1622P}. However, the high-density environment of a collapsing core may affect RGB mass-loss (and HB morphology, in turn) via various channels. Binary stars, for example, are expected to play a major dynamical role in halting core-collapse by releasing energy while becoming more bound \citep[][]{1983ApJ...272L..29H}. This may result in RLOF within binary systems whose primary is an RGB star and whose semiaxis shrinks enough, resulting in enhanced mass-stripping \citep[Binary Channel; ][]{2013A&A...549A.145L, 2013arXiv1305.1996L}. On the other hand, direct stellar collisions and grazing encounters taking place on unbound orbits, are also much more likely to occur when core density increases dramatically due to gravitational collapse, also resulting in enhanced mass-stripping (Collisional Channel; our model presented in this paper). In this context, only a quantitative model may allow us to distinguish between the two channels. Ideally, a quantitative model should be able to predict the HB mass distribution, the rotation rate of HB stars, and their binary fraction. When the comprehensive set of SPH simulations we are running (Moraghan et al. in preparation) is ready, we will be able to predict both mass-loss and angular-momentum deposition paving the way for a quantitative comparison.

\acknowledgments
This paper utilises the HST-snapshot database by the Globular Cluster Group of the Padova Astronomy Department and "The ACS Survey of Galactic Globular Clusters" database publicly available by A. Sarajedini and collaborators. We thank Paolo Ventura for stimulating discussion on stellar evolution models. We wish to thank the referee for comments and questions that helped us clarify several important points. Support for this work was provided by the National Research Foundation of Korea to the Center for Galaxy Evolution Research, and also by the KASI-Yonsei Joint Research Program for the Frontiers of Astronomy and Space Science and the DRC program of Korea Research Council of Fundamental Science and Technology (FY 2012). C. C. acknowledges support from the Research Fellow Program (NRF-2013R1A1A2006053) of the National Research Foundation of Korea. This work received partial financial support by INAF$-$PRIN$/$2010 (PI G. Clementini) and  INAF$-$PRIN$/$2011 (PI M. Marconi).

\appendix
\section{Fitting procedure}
\label{appe}
\begin{figure}
\includegraphics[width=0.8\columnwidth]{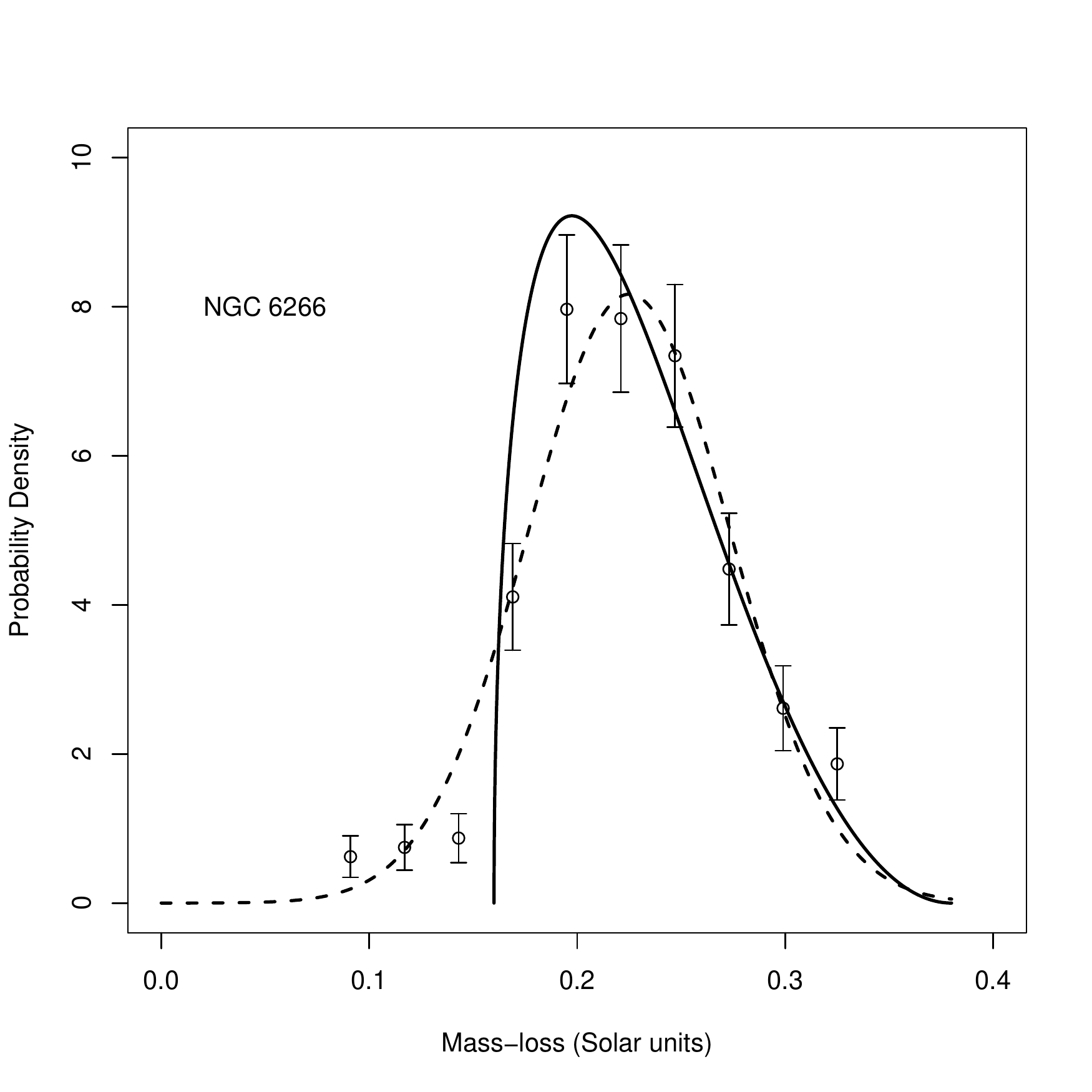}
\caption{Mass-loss distribution in a sample of $309$ HB stars in NGC 6266 from \cite{2002A&A...391..945P} HST data. The points represent the estimated probability density ($y$ axis; based on the normalized counts) for a given mass-loss bin ($x$ axis). The $1$-$\sigma$ counting error is represented by the error bars. The superimposed solid black curve is our best-fit model mass-loss function. {A gaussian with the sample's mean and standard deviation is also shown as a dashed line.}\label{fithist}}
\end{figure}

\begin{figure}
\includegraphics[width=0.8\columnwidth]{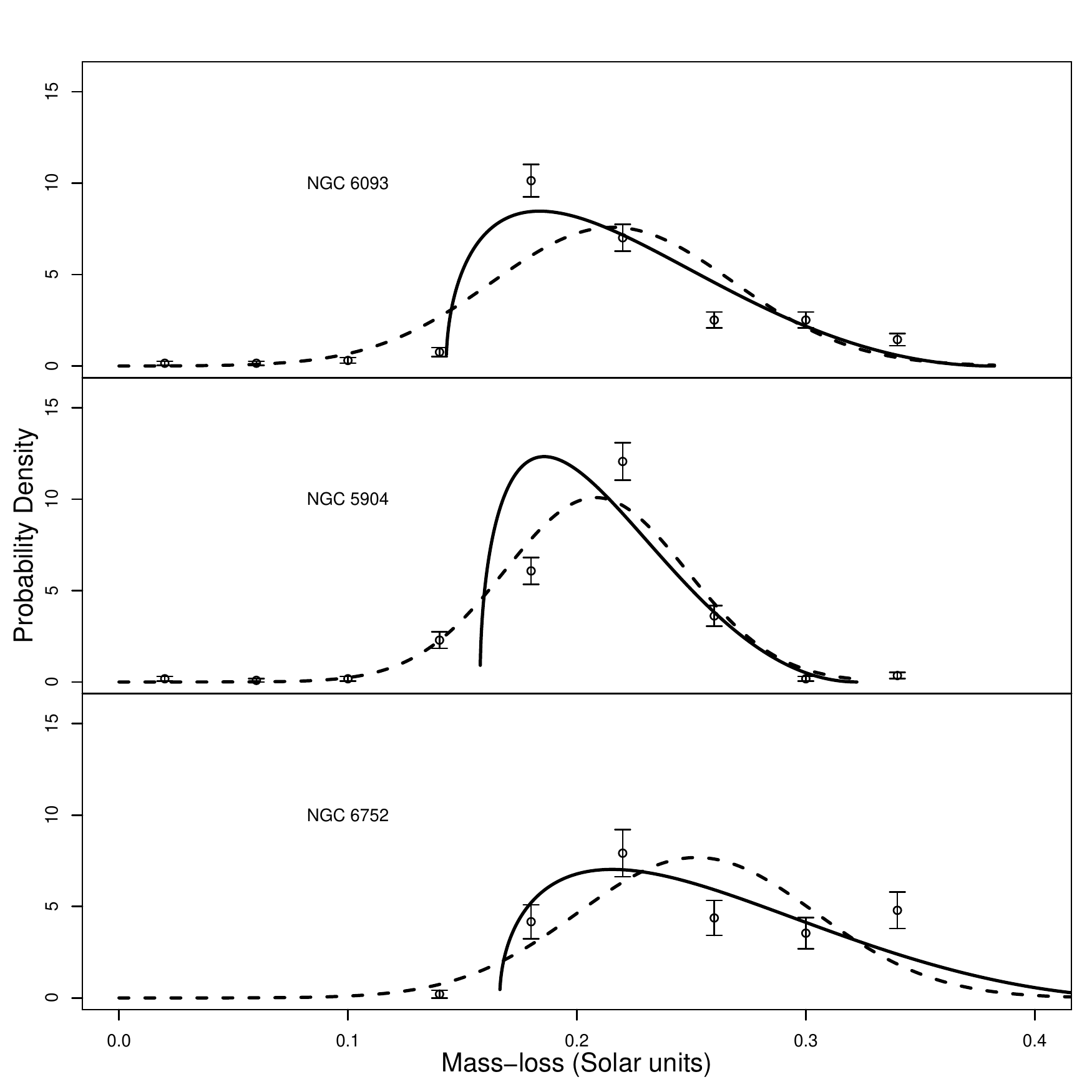}
\caption{Best fit models (solid lines) for mass-loss distributions of NGC 6093, NGC 5904, and NGC 6752 (points). The 1-$\sigma$ counting error is represented by the error bars. {Gaussians with the sample's mean and standard deviation are also shown as dotted lines. Notice how the Gaussian seems to miss the cutoff-like feature at the left side of the plots, but fits NGC 5904 visibly better than our model.}\label{theothers}}
\end{figure}

\begin{figure}
\includegraphics[width=0.8\columnwidth]{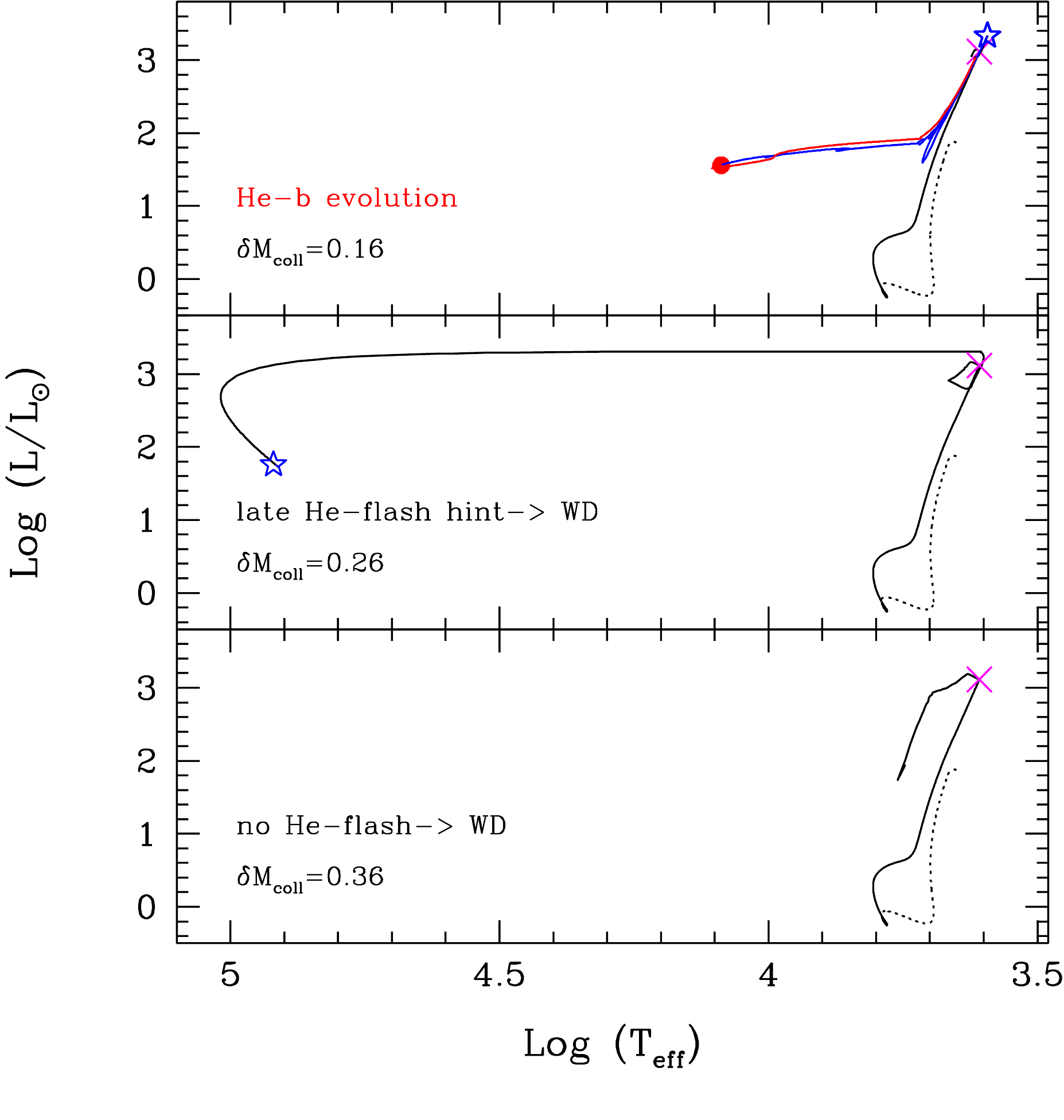}
\caption{The evolution of a star with initial mass $M=0.8 M_\sun$ and chemical composition Z=0.0006, Y=0.24, from the pre-MS (black dotted line) to the RGB and the subsequent phases (solid line). Three representative cases are shown: from top to bottom collision mass-loss is $\delta M_{coll}=0.16 M_\sun$, 0.26 and 0.36. In each panel: the fate of star is labeled; the magenta cross shows the RGB point where the collision mass-loss event takes place; the blue star indicates the first-primary He-flash occurrence. In the upper panel the blue line refers to the evolution toward the ZAHB model (red dot), and the red line illustrates the off-ZAHB evolution.\label{fig:evolution}}
\end{figure}

\begin{figure}
\includegraphics[width=0.99\columnwidth]{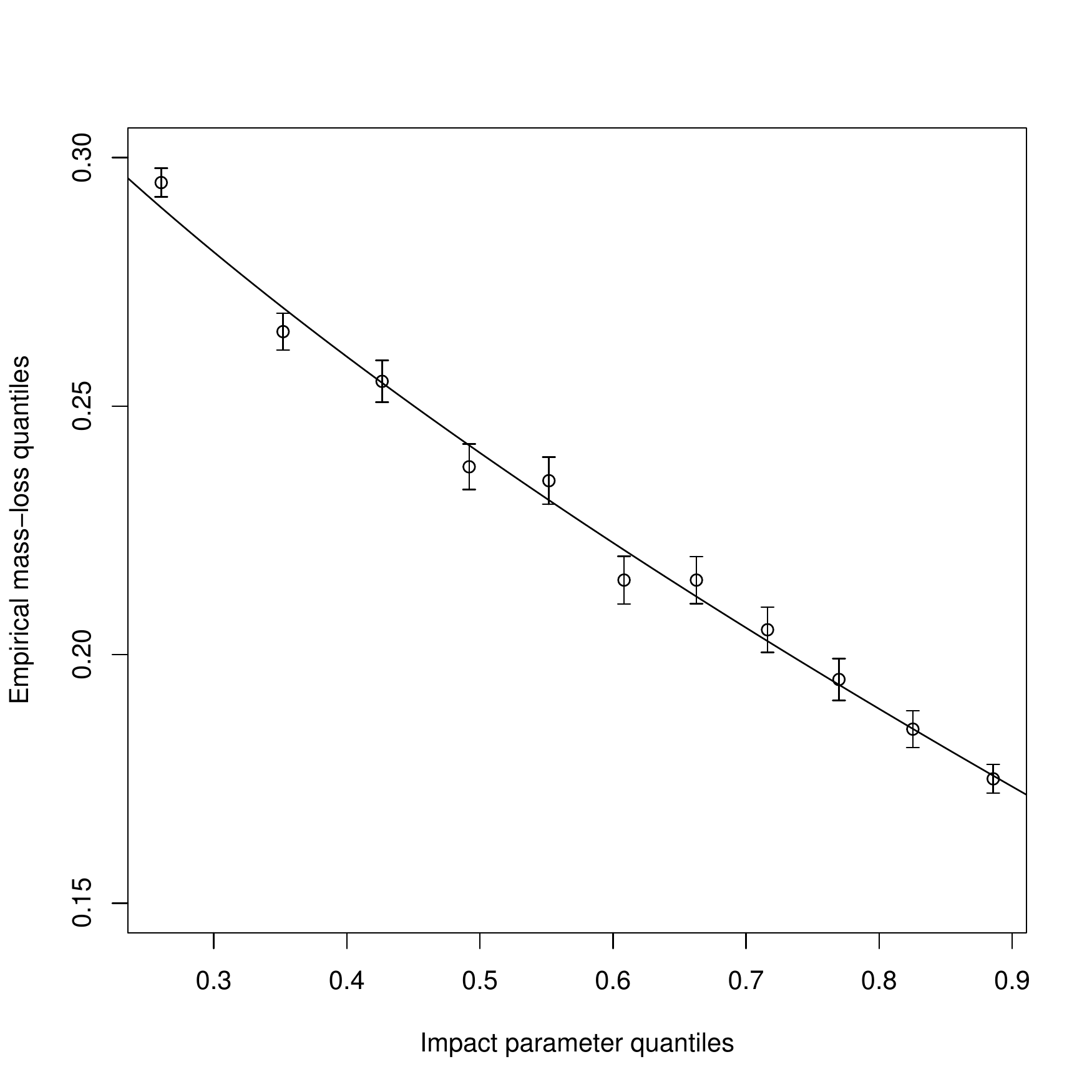}
\caption{Empirical quantiles of mass-loss for the HB stars in NGC 6266 from \cite{2002A&A...391..945P} HST data ($y$ axis), as a function of the corresponding theoretical quantiles of the impact parameter in a Maxwell-Boltzmann distribution ($x$ axis). Error bars are based on the uncertainty of the estimate of the quantiles for a uniform distribution function. The superimposed solid line is our best-fit model.\label{fitquant}}
\end{figure}

Equation~\ref{Hilloverflowbetter} is nonlinear in the $A$ and $B$ parameters. It would in principle be possible to either use a nonlinear fitting procedure to fit it directly to the histograms of mass-loss of each cluster, or to fit it `by eye'. However, it is more convenient to consider Eq.~\ref{HilloverflowVpar}, which is linear in $A$ and $B$. If for every collision that resulted in a mass loss $\delta M_i$ we had the corresponding information on the impact parameter $b_i$, we could directly fit Eq.~\ref{HilloverflowVpar} to the sample of $\delta M_i$ values as a function of $b_i$, using e.g. linear least squares. Unfortunately, we do not know the $b_i$ associated to each $\delta M_i$. However, if we assume that the `real' $\delta M (b)$ relationship is monotonic and the $b$ distribution is actually given by Eq.~\ref{fb}, then the quantiles of the sample $\delta M_i$ can be plotted as a function of the theoretical quantiles of the distribution in Eq.~\ref{fb}. In Fig.~\ref{fitquant} we plot the empirical quantiles of the mass-loss sample as a function of the theoretical quantiles of $b/a$ based on Eq.~\ref{fb} for NGC 6266. This, under the above mentioned hypotheses, amounts to a direct visualization of the function relating mass-loss to the impact parameter, i.e. the `real' $\delta M (b)$ relationship, to which we can then fit Eq.~\ref{HilloverflowVpar} using linear least squares. The values of $A$ and $B$ listed in Table \ref{addita} and the associated errors are obtained in this way. This procedure is not a circular argument: the assumption that $\delta M (b)$ is monotonic and that the underlying distribution of $b$ is actually given by Eq.~\ref{fb} is used only to obtain the best-fitting values of $A$ and $B$, not to validate the model in the first place. Moreover, values independently obtained `by eye' directly on the histograms in Figs.~\ref{theothers} and~\ref{fithist} actually closely match those listed in Table \ref{addita}.

\bibliography{manuscript}
\bibliographystyle{apj}
\clearpage

\end{document}